\documentclass[%
 reprint,
 amsmath,amssymb,
 aps,
]{revtex4-1}

\usepackage{graphicx}
\usepackage{dcolumn}
\usepackage{bm}

\begin{document}


\title{Collective benefits in traffic during mega events via the use of information technologies}

\author{Yanyan Xu}
\affiliation{%
 Department of Civil and Environmental Engineering, MIT, Cambridge, MA 02139, USA
}%


\author{Marta C. Gonz{\'a}lez}
\altaffiliation[]{martag@mit.edu}
\affiliation{%
 Department of Civil and Environmental Engineering, MIT, Cambridge, MA 02139, USA
}%
\affiliation{
 Engineering Systems Division, MIT, Cambridge, MA 02139, USA
}%

\date{\today}

\begin{abstract}
 Information technologies today can inform each of us about the best alternatives for shortest paths from origins to destinations, but they do not contain incentives or alternatives that manage the information efficiently to get collective benefits. To obtain such benefits, we need to have not only good estimates of how the traffic is formed but also to have target strategies to reduce enough vehicles from the best possible roads in a feasible way. Moreover, to reach the target vehicle reduction is not trivial, it requires individual sacrifices such as some drivers taking alternative routes, shifts in departure times or even changes in modes of transportation. The opportunity is that during large events (Carnivals, Festivals, Sports events, etc.) the traffic inconveniences in large cities are unusually high, yet temporary, and the entire population may be more willing to adopt collective recommendations for social good. In this paper, we integrate for the first time big data resources to quantify the impact of events and propose target strategies for collective good at urban scale. In the context of the Olympic Games in Rio de Janeiro, we first predict the expected increase in traffic. To that end, we integrate data from: mobile phones, Airbnb, Waze, and transit information, with the unique information about schedules, location of venues and the expected audience for each match. Next, we evaluate the impact of the Olympic Games to the travel of commuters, and propose different route choice scenarios during the morning and evening peak hours. Moreover, we gather information on the trips that contribute the most to the global congestion and that could be redirected from vehicles to transit. Interestingly, we show that (i) following new route alternatives during the event with individual shortest path (selfish strategy) can save more collective travel time than keeping the routine routes (habit strategy), uncovering the positive value of information technologies during events; (ii) with only a small proportion of people selected  from specific areas switching from driving to public transport (mode change strategy), the collective travel time can be reduced to a great extent. Results are presented on-line for the evaluation of the public and policy makers (\protect\url{www.flows-rio2016.com}). 
\end{abstract}

\maketitle


\section{\label{sec:intro}Introduction}

There is complex relationship between transportation, land use and the urban form. Technological innovations, socio-demographic shifts and political decisions shape the way people move in cities~\cite{bettencourt2007growth, gonzalez2008understanding, lederbogen2011city, batty2012smart, batty2013theory, bettencourt2013origins}. The amount of time invested to move to work every day \cite{song2005self, hasan2013understanding, jiang2013review, pappalardo2015returners} has important implications in the well functioning of our cities. They affect total energy use, equity, air pollution, and urban sprawling.  Given this impact, master plans of urban transportation need to be both technically sound and politically feasible~\cite{crucitti2006centrality, dye2008health, scellato2010traffic, setton2011impact, knittel2011caution, colak2016understanding}. This becomes a more pressing need when preparing for large events that unusually stress further the use of the available infrastructures and put at risk the overall success of the event if the planning is poor. 

 In their best attempts, goals of an urban transportation plan seek to: (a) avoid long and unnecessary motorized travel, (b) shift the movement of people to socially efficient modes such as walking, biking, and public transit, and (c) improve the technology and operational management of transportation services.  To reach these goals plans today try to promote the use of bus rapid transit (BRT), congestion charging, or bike-sharing. But much less is done to develop real time information platforms that provide the value of choices for the social good. Nowadays, the most popular information platforms such as Waze or Google Transit Feeds give us individual information about travel times but do not take into account global information to better govern the system. One limitation may be that the main set of tools and skills in  urban transportation planning were developed in the seventies, before the information age, and relied on the results of travel diaries, from which estimating urban travel demand could be a very time consuming task. Second, even if we could estimate the best collective solutions, urban transportation faces the ``tragedy of the commons''. Meaning that streets are a shared-resource system where individual users act independently according to their own self-interest behaving contrary to the common good of all users by depleting that resource through their collective action.  However, there may be instances, when the population may be more prone to take actions for collective benefits.

This may be the case during large urban events, which gather people from different places. These represent important changes in routine activities of the population. Large-scale events happen every year around the world, such as Olympic games, world expositions, concerts, pilgrimage, etc. These popular events attract massive of participants or tourists traveling to one destination, thereby produce great opportunities also huge pressures to transportation and the environment, especially for  cities with high population~\cite{burgan1992economic, dork2010visual}. Past research has tried to estimate the impact of events to the economy and air quality of the host city~\cite{hotchkiss2003impact, lee2005critical, peel2010impact}. In the context of traffic management during large-scale events, previous efforts focus on ensuring the travel of participants efficient and unimpeded. However, the disturbances to the travel of the local population are not taken into account. Traditionally, the government seeks to reduce motorized travels globally by ending plate number, but without making smarter use of the travel information~\cite{levinson2003value, selten2007commuters}.

In this work, we evaluate the impact of large-scale events to the traffic in the host city and strategies to overcome it. Especially, we aim at understanding the change of travel demand when large-scale events come, and addressing reasonable demand management strategy to mitigate the traffic congestion during events. We take the Summer Olympics 2016 as an example to study the impact of large-scale events to the travel of local population. The Summer Olympics will be held in Rio de Janeiro, which is one of the most congested cities in the world according to the TomTom's report on global traffic congestion~\cite{TomTom2016}. A study released by the Industry Federation of the State of Rio de Janeiro (FIRJAN) confirms that traffic congestion has tremendous economic costs as well. The study found that congestion costed the cities of Rio and S\~{a}o Paulo roughly USD 43 billion in 2013 alone. The loss amounts to about $8\%$ of each metropolitan area's Gross Domestic Product (GDP). This is greater than the estimated budget for transport capital investment in Brazil, Mexico, and Argentina combined. 
Traffic congestion originates from the imbalanced development of travel demand of vehicles and the road network supply~\cite{braess2005paradox, Black2010understanding}. For a booming city, the traffic congestion can be mitigated through constructing more roadways and transit infrastructure. But for mature urban areas like Rio, opportunities for further investments in transportation infrastructure are often limited~\cite{luten2004mitigating}. 

As TomTom reported, the commuters in Rio spend nearly $70\%$ extra time during the peak hours in 2015. The opening of Olympics will undoubtedly aggravate the travel delay of the local population as the increment of demand and reduction of supply. The International Olympic Committee (IOC) predicts $0.48$ million tourists in Rio for Olympics, which is about $7.5\%$ of Rio population. To understand the impact of Olympics, we estimate the person and vehicle travel demand of local population using mobile phone data, also known as call detail records (CDRs) combined with Waze data (see SI for the description). The travel time of commuters are estimated during the morning and evening peak hours and compared with Google maps in the same hour. During the Olympics, we predict the origin and destination of tourists using the Olympic Games' schedule, information of venues, Airbnb properties and hotels, and then split the taxi demand of tourists out from the total demand in each hour. Afterward, the taxi demand together with the local vehicle demand are assigned to the road network under three routing scenarios: \emph{habit}, \emph{selfish}, and \emph{altruism}. We then predict the travel time of tourists and evaluate the increment of local commuters' travel time under these routing scenarios in the peak hours. In addition, to mitigate the traffic congestion within a short time, we propose a reasonable and easily reached \emph{mode change} strategy, which in essence target the best fraction of travelers that could change from driving to metro and BRT. To this end, we uncover the origin destination pairs with most contribution to the collective travel time and  consider the overall benefit of taking one vehicle out of that pair. Finally, we demonstrate the effectiveness of the proposed demand management strategy by comparing with a benchmark program that reduces the same number of vehicles randomly distributed.

\section{\label{sec:resul}Results}

\subsection{\label{sec:OD}Travel Demand Estimation}

\vspace{3mm}
\noindent \textbf{Travel demand estimation before the Olympics.} 
In previous studies, we have estimated the average hourly travel demand successfully using CDRs from mobile phones (include the timestamp and location for every phone call or SMS of innominate users), census records, and surveys data in Rio de Janeiro~\cite{colak2015analyzing, toole2015path, alexander2015origin, colak2016understanding}. In the travel demand estimation framework, the stay locations of each user are recognized and labeled as home, work, or other. Consequently, we classify the trips of each person into three cases: home-based-work (commuting), home-based-other, and non-home-based. After aggregating the trips from CDRs and the census data, we could get a reasonable origin-destination (OD) matrix with different travel purposes. Finally, the estimated person (vehicle) demands are $1.69$ ($0.44$) million and $1.61$ ($0.41$) million during the morning and evening peak hour on weekdays in Rio municipality, respectively (see Supplementary Figure 1 and Note 1). Afterward, we can extend the vehicle demand to variations of five weekdays, with the help  of records from Waze Mobile~\cite{Waze2015}. Waze provided the records of wazers for $7$ months in 2015. The data sets include the location of user, timestamp, level and duration of jam, average speed, and length of the queue, etc. We relate the average length of the queue in the road network (estimated with Waze data) as proportional to the total vehicle demand in this hour (estimated with mobile phone data). Consequently, we extend the 24-hour demand estimated by CDRs to 5 weekdays according to the different average queue length on different weekdays (see Supplementary Note 2).

\begin{figure*}[htbp]
\centering
\includegraphics[width=0.96\linewidth]{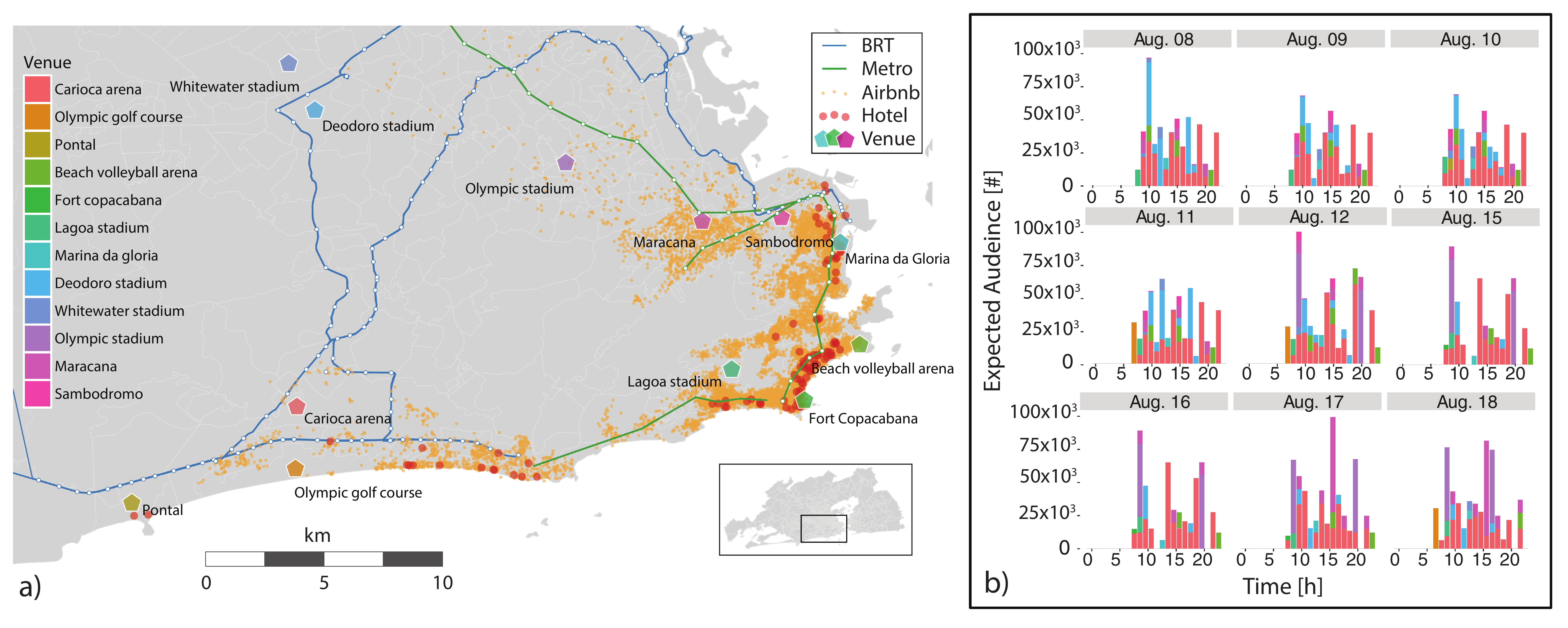}
\caption{\textbf{Locations of venues, tourists' residences and number of spectators per venue per hour.} (a) The locations of $12$ Olympic venues, the Metro and BRT lines in Rio, the locations of hotels and distribution of Airbnb properties. Most venues are near to Metro/BRT stations, as well as hotels and most Airbnb properties. We distribute tourists around the $13,400$ Airbnb properties and $106$ hotels. Metro and BRT will likely be the first choice for most spectators. (b) The number of spectators arriving at each venue per hour on $9$ weekdays during Olympic Games. The largest indoor stadium, Carioca arena is also the busiest one.}
\label{fig:map}
\end{figure*}

\vspace{3mm}
\noindent \textbf{Travel demand estimation during Olympics.}
To build the OD matrix during Olympics, the two issues to be addressed are the new locations of origins and destinations and the flow between them. To facilitate this, we only consider the flow from spectators' residences to venues. Figure~\ref{fig:map}a represents the location of $12$ Olympic venues, the distribution of Airbnb properties and hotels, the metro and the BRT line in Rio. Most of the tourists' residences are distributed in the southeast coastal area. As planned by the municipal government, most venues are located around the metro or BRT stations, which makes public transportation quite convenient most of the time. 

The person travel demand equals to the sum of local demand before Olympics and the number of people going to stadiums from their residences in the same time interval. To achieve this, the number of spectators arriving each venue is estimated hourly based on the Olympic game schedule and capacities of venues. Figure~\ref{fig:map}b shows the results on weekdays during Olympics. The maximum number of spectators is nearly $0.1$ million, which is considerable in contrast to the number of commuters in peak hour. To determine how many spectators depart from hotels to venues and when, we make the following assumptions: (i) $30\%$ spectators will departure $1$ hour ahead; $40\%$ spectators will departure $2$ hours ahead; the others will departure $3$ hours ahead. (ii) Despite a part of spectators are from local, as the distribution of Airbnb properties is similar to the distribution of local population, consequently, we consider all of the spectators are from Airbnb properties and hotels, and name them as tourists in the rest of the paper. To estimate the origin of all tourists per hour, we assign all of the spectators to Airbnb properties and hotels according to their capacities (see Supplementary Note 3). 

\begin{figure*}[htbp]
\centering
\includegraphics[width=0.96\linewidth]{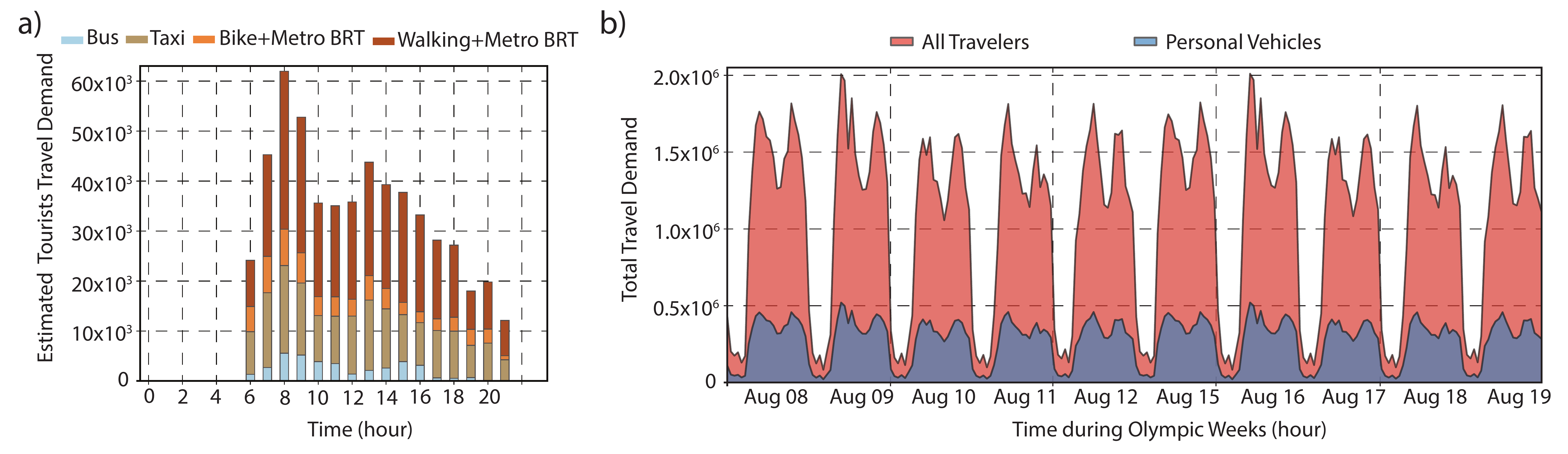}
\caption{\textbf{Estimation of tourists' travel mode and travel demand during Olympics.} (a) Results of Travel mode split for tourists on August 8.  (b) The predicted  travel demand by total  number of people and vehicles during Olympics on $10$ weekdays from August 08 to 19. The person demand estimates are the aggregation of local travelers and tourists going to venue from hotel in the same hour. The vehicle demand is the aggregation of local vehicle demand and taxi used by tourists.}
\label{fig:demand}
\end{figure*}

To estimate the additional vehicle demand during Olympics, we estimate the travel mode of tourists in each hour, distributing them among public transportation or taxi. Specifically, we split the travel mode of tourists into 4 categories, e.g. walking and Metro/BRT, bike and Metro/BRT, taxi, and bus, depending upon the following features: distance to metro/BRT stations, travel time, the number of mode transitions. Figure~\ref{fig:demand}a shows the results of travels by mode on August 8 (Monday). As expected, most tourists choose Metro/BRT because both of their hotels and venue are near to Metro/BRT stations. Nonetheless, during the daytime, we estimate that about $10,000$ tourists will choose taxi to the venue per hour, which will produce considerable additional vehicle demand compared with before the Olympics (see Supplementary Figure 2).

Figure~\ref{fig:demand}b shows the total person and vehicle demand on $10$ weekdays from August 8 to 19. The vehicle demand equals to the local vehicle demand from CDRs plus the taxi flow by tourists. The morning peak is around 9:00 and the evening peak is around 18:00. During the peak hour, about $27\%$ of the population are traveling, the number increase about $60,000$ during Olympics. Consequently, traffic in the city will be more congested, especially for the path from tourists' residences to venues.

\subsection{\label{impact}Travel time estimation and impact analysis}

Before Olympics, we assign the drivers to the routes with the shortest paths. This is a common approximation to the complex problem of route selection, in which  the travel routes and times are estimated by assigning the local demand to the road network using an assignment model - the user equilibrium (UE) model, which means no driver can unilaterally reduce his/her travel time by changing routes. In our implementation of UE model, the travel times of  links depend on the volume-over-capacity ratio (VoC) with the Bureau of Public Roads (BPR) function:
\begin{equation}
\label{equ:bpr}
t_e(v_e) = f_s \left[1 + \alpha \left( \frac{v_e}{C_e}\right)^\beta \right] \times t_e^f
\end{equation}
where $t_e(v_e)$ is the average travel time on link $e$; $t_e^f$ is the free flow travel time on this link; $f_s$ is a scale factor and not less than $1$. The coefficients in BPR are calibrated using field data collected by surveillance cameras as $f_s=1.15$, $\alpha=0.18$, $\beta=5.0$. We compared our estimated travel times of top commuter OD pairs with Google map in the same hour and found good agreement (see Supplementary Figure 1, Figure 3, and Note 4).

The Olympics will disturb the routes of a fraction of travelers, especially those with routine routes congested by the trips to the games and the reduced capacity of the Olympic lanes. Olympic lanes will be dedicated to athletes by separating a lane from some roads, while tourists and local travelers can not use the reserved lane. In our calculations we take into account this new reduced capacity.
We explore the effects of three distinct behavioral choices: (i) \emph{habit}: All travelers will follow their routine travel routes even if this route is having more congestion during the Olympics; (ii) \emph{selfish}: travelers have good knowledge of the traffic situation each of them will choose the route with shortest travel time, which follows the UE model; (iii) \emph{altruism}: travelers are follow the travel routes for the best case scenario for the total travel time. In this case, the travel route of each traveler is assigned via system optimization. Under these three scenarios, the traffic states on the roads are diverse under the three scenarios (see Supplementary Figure 4).

\begin{figure*}[htbp]
\centering
\includegraphics[width=0.96\linewidth]{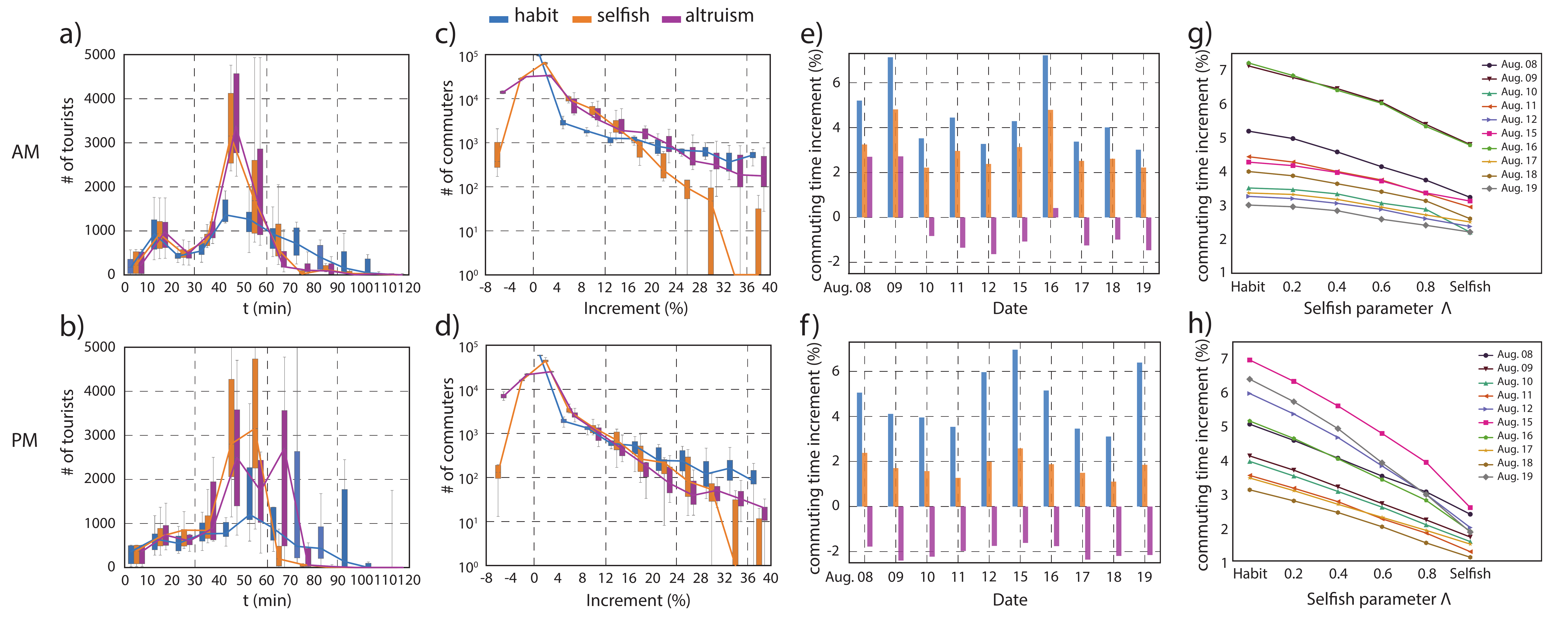}
\caption{\textbf{Tourists' travel times and impact to local commuters under three scenarios.} (a-b) Box plot of tourists' taxi travel time during the morning peak on $10$ weekdays. In morning peak hour, the average tourists travel time of \emph{habit}, \emph{selfish}, and \emph{altruism} are 66 mins, 44 mins, and 43 mins respectively. In evening peak hour, they are 62 mins, 43 mins, and 47 mins respectively. (c-d) Box plot of commuters' travel time percentage increment on $10$ weekdays. The number of commuters is scaled with $\log$ function. Negative percentage increments indicate people could reach smaller travel time than before Olympics. (e-f) Average commuting time percentage increment comparison of three scenarios of $10$ weekdays in the morning and evening peak hour, respectively. (g-h) The change of average percentage increment with the increase of selfish parameter $\Lambda$ on each weekday.}
\label{fig:impact}
\end{figure*}

To access the impact of different routing strategies during the Olympics, we predict the travel time of tourists and local commuters under the three ideal scenarios. Figure~\ref{fig:impact}a and~\ref{fig:impact}b illustrate the box plot of distribution of tourists' travel time during the morning and evening peak hour on $10$ weekdays, respectively. The \emph{habit} scenario always perform worse than \emph{selfish} and \emph{altruism} as local travelers will not give their way to tourists even they are suffering more serious congestion and could find shorter path. \emph{Selfish} and \emph{altruism} scenarios, by contrast, allow travelers choose their route toward their own or others' benefit. To evaluate the impact of Olympics to local commuters, we calculate the average percentage increment of commuter's travel time as
\begin{equation}
\label{equ:increment}
I_{comm} = \frac{\sum_{od \in OD}{(t_{od}^{Olym} - t_{od}^{before})f_{od}^c}}{\sum_{od \in OD}{t_{od}^{before} f_{od}^c}} \times 100\%
\end{equation}
where $od$ is one of all the $OD$ pairs; $f_{od}^c$ refers to the number of commuters; $t_{od}^{Olym}$ and $t_{od}^{before}$ refer to the travel time in the $od$ route before and during Olympics, respectively. $I_{comm}$ can be negative as \emph{selfish} or \emph{altruism} allows some commuters find a shorter path than before. Figure~\ref{fig:impact}c and ~\ref{fig:impact}d depict the distribution of commuters travel time in a log scale on weekdays. More people have larger travel times ($I_{comm} > 20\%$) under the \emph{habit} scenario than under \emph{selfish} or \emph{altruism} scenarios. Moreover, in contrast with \emph{selfish}, 
\emph{altruism} raises the number of commuters suffering high percentage increment but earns remarkable benefits for more commuters.
This is because of the essence of \emph{altruism}: while a small fraction of people sacrifice with longer travel times via detour to less popular routes~\cite{colak2016understanding}, the overall saving in travel time is larger. We find interesting contrasts to studies made during commuting
routines in Rio~\cite{colak2016understanding}, here we see that the difference in travel-times between the \emph{altruism} and \emph{selfish} are smaller, but both are in strong contrast to the \emph{habit} scenario. This shows the clear benefits of information technologies to help decrease congestion during the events.

Figure~\ref{fig:impact}e and ~\ref{fig:impact}f illustrate the average percentage increment per day. The percentage increment of the \emph{habit} scenario is always larger, followed by \emph{selfish} and \emph{altruism}. Furthermore, certain peak hours are subject to most serious delay, e.g., morning peaks on August 09 and 16, evening peaks on August 12, 15, and 19.

However, in practice, the most plausible travel behaviors of travelers is between \emph{habit} and \emph{selfish}, which means a fraction of people change path toward shortest travel time, others keep their routine routes. To examine such intermediate state, we define a selfish parameter $\Lambda$ signifies the fraction of selfish travelers.
$\Lambda$ ranges from $0$ to $1$, where $0$ equals to the \emph{habit} scenario, and $1$ equals to the \emph{selfish} scenario. Specifically, the travelers in each OD pair seek their shortest travel time with a percentage of $\Lambda$ and their routes need to be reassigned, others are following their habit routes. For each link, it can be occupied by habit flow and selfish flow. The habit flow is calculated as $v_e^{habit} \cdot (1-\Lambda)$, where $v_e^{habit}$ is the link volume under \emph{habit} scenario. The selfish flow $v_e^{selfish}$ is obtained by assigning the selfish demand using UE model. Therefore, the VoC is calculated by:
\begin{equation}
\label{equ:selfish}
 VoC_{e} = \frac{v_e^{habit}(1-\Lambda) + v_e^{selfish}}{C_e}
\end{equation}
and the BPR function in Equation~\ref{equ:bpr} is used to estimate the travel time on each link. For each OD trip, the total commuting time also contains two parts: $(1-\Lambda) \cdot f_{od}^c \cdot t_{od}^{habit}$ and $\Lambda \cdot f_{od}^c \cdot t_{od}^{selfish}$, where $t_{od}^{habit}$ is the travel time under \emph{habit} scenario and $t_{od}^{selfish}$ is the shortest travel time under selfish parameter $\Lambda$.
Figure~\ref{fig:impact}g and~\ref{fig:impact}h indicate the average increment for commuters on each weekday with different selfish parameters. The increment percentage decreases with the increase of $\Lambda$, which indicates that the impact of Olympics reduces if more travelers are selfishly looking for their best route as opposed to using their routine routes. 

Most of the transportation planning strategies designed to reduce motorized vehicles are applied independently of origin and destination of the travelers, in consequence they are very costly in terms of the percentage of car reduction (usually $10\%$ of the cars selected by ending digit in the plates), to achieve very modest benefits in travel times, usually of the order of $2\%$~\cite{wang2012understanding}.
Based on the estimation of travel delays of commuters under the \emph{selfish} scenario, we explore the spatial impact of Olympics to commuters in relation to their living and working places. To achieve this, we average the percentage increment of commuter trips to origin and destination zones. Results indicates commuters who live in the northeast of Rio suffer serious impact in the morning peak hour (see Supplementary Figure 5). In addition, people working in the eastern coastal area suffer travel delays the most in both of the morning and evening peak hour. We also find the densely populated Governador Island always suffer critical delay as one of the two bridges between the island and mainland are set as Olympic lanes. 

\begin{figure*}[htbp]
\centering
\includegraphics[width=0.90\linewidth]{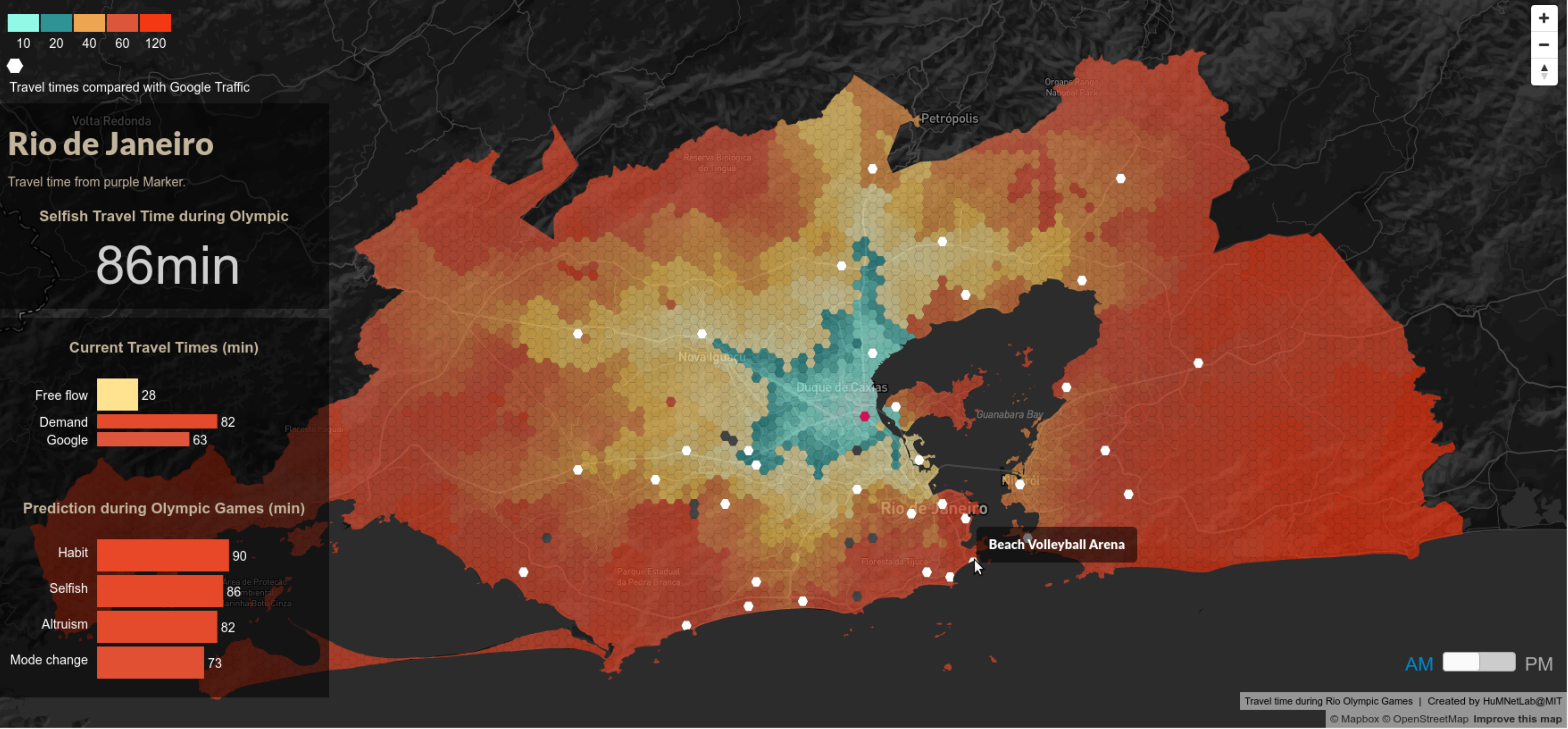}
\caption{\textbf{Interactive visualization of travel times before the Olympics and during the Olympics via various strategies of mobility.} The purple hexagon reflects the origin of trips. The white hexagons are associated with the Google travel time for comparison. The colors of other hexagons reflect the travel time from origin to them.
Results are presented on-line for the evaluation of the public and policy makers, see \protect\url{www.flows-rio2016.com}.}
\label{fig:platform}
\end{figure*}

To facilitate the policy making, we visualize the travel time before and during Olympics all over the metropolitan area of Rio, as shown in Figure~\ref{fig:platform}. From the visualization, travelers can explore their travel time increment by Olympics during the peak hours. Besides, the platform provides the travel time under different scenarios, which helps the travelers and policy makers realize the collective benefits generated by the travel demand management strategy.

\subsection{\label{strategy} Informed mode change strategy}

Aiming to mitigate the traffic congestion during Olympics, the government of Rio de Janeiro has made great efforts, such as enhancing the capacity of the traffic network, extension of the public transportation infrastructure, including construction of new metro and BRT lines. However, considering the massive economic and time cost of construction, in this work we propose an efficient strategy to manage the travel demand with the present transport infrastructure, concretely, reducing a fraction of vehicle demand toward relieving congestion  during the peak period to the most extent. 

With the purpose of selecting the critical trips to reduce, we quantitatively evaluate the contribution of each OD trip to the collective travel time. Namely, we consider the following question: how much time will the collective save if we take one vehicle out from the existing demand? We represent the road network as a directed acyclic graph $\mathcal{G}(\mathcal{N}, \mathcal{E})$, where $\mathcal{N}$ is the set of nodes, and $\mathcal{E}$ is the set of directed edges. After assigning the travel demand to the road network, each road segment $e \in \mathcal{E}$ is associated with volume $v_e$ and travel time during traffic $t_e$. First, for a road segment, we estimate the travel time saving of others if we reduce one vehicle using the marginal edge cost, which is the partial gradient of total travel time over the current volume. For each edge, we have:
\begin{equation}
\label{equ:linkcost}
\begin{aligned}
MC_e &= \frac{\partial (v_e t_e)}{\partial v_e} \\
 &= t_e(v_e) + f_s \alpha \beta \left( \frac{v_e}{C_e}\right)^\beta \times t_e^f
\end{aligned}
\end{equation}
where edge travel time $t_e$ is calculated using the calibrated BPR function in Equation~\ref{equ:bpr}. The marginal edge cost $MC_e$ consists of two terms: the first one $t_e$ reflects the travel time of one vehicle and the second would be the saved travel time by other vehicles in the same edge. The travel route $p_{i,j}$ of each OD trip $(i,j)$ is the set of edges on the path. Consequently, we calculate the marginal path cost of OD pair $(i,j)$ as the sum of $MC_e$ for the edges traversed by the path:
\begin{equation}
\label{equ:pathcost}
MC_p = \sum_{e\in \mathcal{E}}\delta_{ep} {MC_e}
\end{equation}
where $\delta_{ep}$ is the delta function, which is $1$ if edge $e$ is traversed by path $p$, $0$ otherwise.

Larger values of $MC_p$ indicate more collective travel time would be saved if we take the trip out. Consequently, a sensible strategy is to reduce the demand from top-ranked OD pairs. To formulate a feasible strategy, we only consider the trips which origins and destinations are both nearby the metro or BRT stations, which means these people could switch to public transport other than driving. In our experiments, we examine the maximum distance from centroid of zone to nearest station as $1km$, $2km$, $3km$, respectively. First, we select the OD trips in certain distance to the nearest metro or BRT station. Then, calculate the $MC_p$ for each trip and reduce $60\%$ demand from the top-ranked trips. The number of top-ranked trips ranges from $1,000$ to $10,000$. Finally, we reassign the remainder demand to the road network and check the reduction of the collective travel time. 
As a benchmark, we keep the same number of total reduced trips but uniformly distribute them to all OD pairs near Metro and BRT stations.

\begin{figure*}[htbp]
\centering
\includegraphics[width=0.80\linewidth]{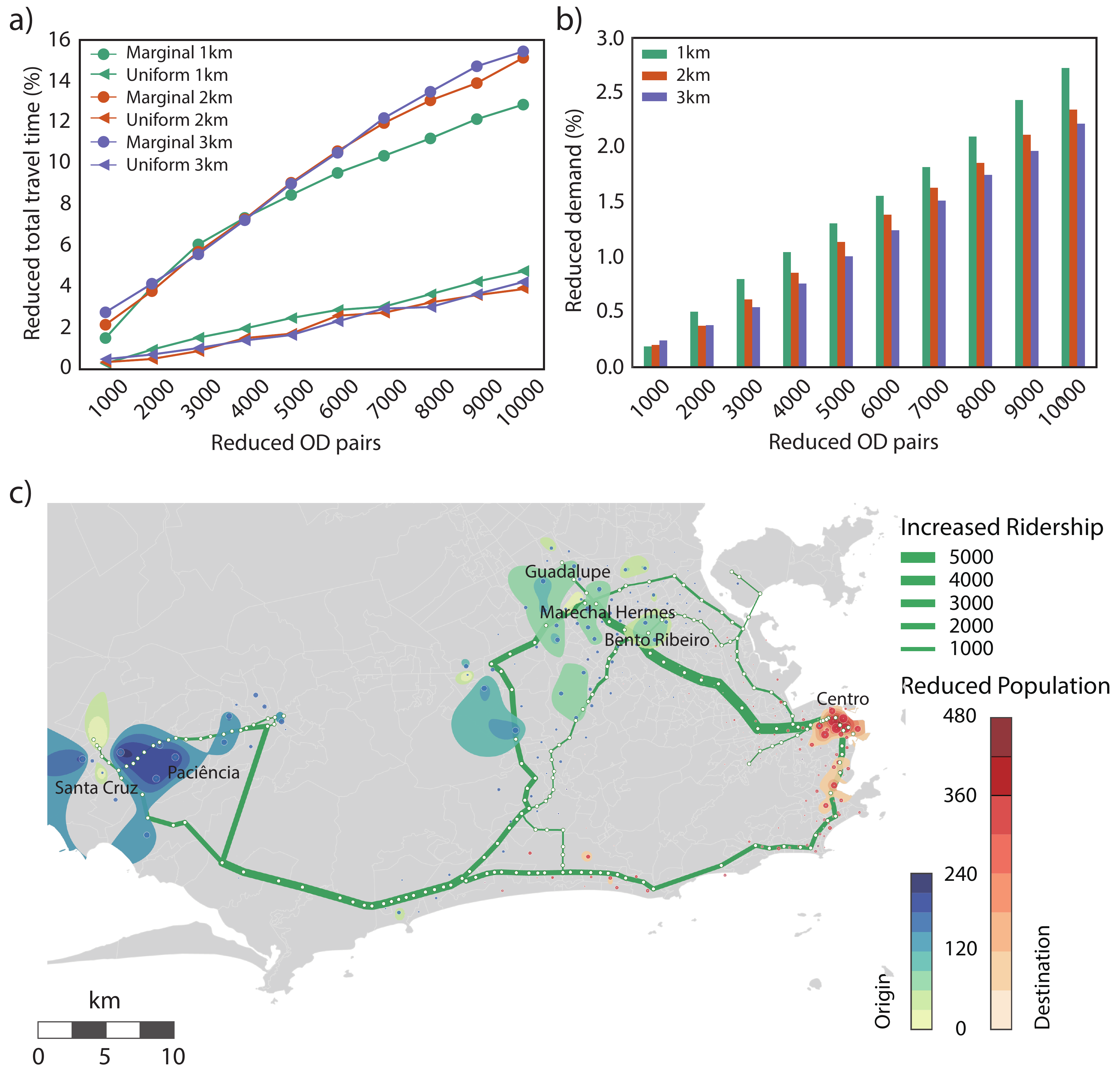}
\caption{\textbf{Demand management results during the morning peak hour.} (a) Collective travel time reduction under different strategies. The reduction of collective travel time rise linearly over the number of reduced OD pairs. The slope of marginal strategy equals to $1.45\times 10^{-3}$, $1.64\times 10^{-3}$, and $1.67\times 10^{-3}$ for $1km$, $2km$, and $3km$, respectively; The slope of uniform benchmark equals to $0.47\times 10^{-3}$, $0.39 \times 10^{-3}$, and $0.40\times 10^{-3}$ for $1km$, $2km$, and $3km$, respectively. As can be seen, marginal cost based strategy reduces the collective travel time with more than $3$ times of uniform based strategy. (b) The reduced demand under different strategies. In general, a greater extent produces lower demand to reduce. With the increase of the reduced OD pairs, the difference of total reduced demand is more evident. (c) Addition ridership to Metro/BRT line and reduced population around stations with configuration \{$2km$, $6000$ OD pairs\}. The width of the metro and BRT line reflects the increased ridership by strategy. Blue and red in different area reflect the origin and destination of reduced demand, respectively. The deeper the color, the more the people need to switch to metro or BRT from driving.}
\label{fig:strategy}
\end{figure*}

Figure~\ref{fig:strategy}a illustrates the reduction of collective travel times as a percentage of the travel time before the strategy, which approximately follows a linear relationship with the number of reduced OD pairs. Interestingly, in contrast to the uniform benchmark, the strategy based on marginal costs can be more effective by a factor of five. For example, if we reduce $60\%$ flow from the selected $5000$ OD pairs at the range of $2km$, this represents $1.14\%$ of the total flow. In that case, the reduction in the percentage of collective travel time is more than $10\%$ with the marginal cost strategy and only $2\%$ with the uniform benchmark case. In addition, whether for marginal cost strategies or uniform benchmark, different distances produce similar results for the same number of OD pairs. However, as shown in Figure~\ref{fig:strategy}b, greater extent indicates a lower demand needed to reduce. The reason is that a greater extent provides more candidates, which leads us to choose the OD pairs with more contribution to the collective travel time saving. 

Figure~\ref{fig:strategy}c presents the spatial distribution of the reduced demand with strategy \{$2km$, $6000$ OD pairs\}. The strategy reduces the collective travel time by $10.6\%$ at the expense of $1.4\%$ decrease of the demand, and improve the average speed of all vehicles from $37.08 km/h$ to $39.94 km/h$. Through the strategy, a fraction of local commuters' and tourists' travel time are shorten, especially for the travelers with a long trip (see Supplementary Figure 6). Interestingly, the distribution of destinations concentrates a very small area, Centro of Rio. Meanwhile, the distribution of origins concentrates two areas, the west end of BRT line and the west end of metro line. Proposed strategy suggests people live in two neighborhoods in the West Zone of Rio (e.g. Santa Cruz and Paci\^encia) and three neighborhoods in the North Zone of Rio (e.g. Guadalupe, Marechal Hermes, and Bento Ribeiro) switch from driving to BRT or metro line during the morning peak hour, especially who work in Centro of Rio. Moreover, Figure~\ref{fig:strategy}c gives the additional ridership produced by the proposed demand management for each segment of the metro and BRT line. As can be seen, the maximum increase is $5,000$ travelers in the morning peak hour, which is acceptable in contrast with the capacity of metro and BRT, $\sim 30,000$ passengers per hour per direction.

\section{\label{discuss}Discussion}

Mega events can greatly benefit the host city in many aspects, such as attracting investment, stimulating economy, and attracting countless tourists, etc. Nevertheless, it also exerts disruptions in the routine of the city. One of the most feared costs 
by the population is the increase in travel times, especially for already dense cities, which are more likely to host the event. In the run-up to Olympics, city planners need estimates on how the traffic will be affected, in order to establish proper policies to cope with the impact. However, the current impact evaluation on travelers is mostly confined to qualitative studies with anecdotal experience of events management, but we lack of  quantitative methods to support the strategies. Mostly due to difficulties of data availability to estimate travel demand. In this work, we present a method to estimate urban travel demand and the time increments to commuters during a large event by integrating multiple and large scale data resources. Moreover, we propose reasonable and effective strategies to help mitigate the increase in congestion.

As a case of study, we take the 2016 Summer Olympics in Rio de Janeiro. The large inflow of tourists increases the travel demand while the establishment of Olympic lanes decreases the road network supply. The main task is to estimate the raise in the demand-to-supply ratio and how this will affect travel times. Fist, we estimate the person and vehicle travel demand during Olympics in Rio by the prediction of the number of tourists in traffic and their travel mode. In particular, we can already expect greater number of tourists are traffic during the morning peaks of August 8th, 12th, and 15th, as well as the evening peaks of August 12th, 15th, 16th, and 17th. By estimating the routing of travelers under three distinct scenarios, \emph{habit}, \emph{selfish}, and \emph{altruism}, we assess quantitatively the impact of Olympics to commuters. We find that the \emph{habit} scenario produces the greatest travel times, followed by \emph{selfish} and \emph{altruism}. For some peak hours, the increment in the percentage of travel times of all commuters can be up to $7\%$ if people follow their routine routes. The \emph{selfish} scenario, which is the maximum benefit possible just changing routes, still produces about $5\%$ of the increment for the most affected peak hours. This is in agreement with the magnitude of savings reported by c\c{C}olak \emph{et al.}~\cite{colak2016understanding} in routine conditions. They showed that the collective travel times could be decreased at most by $4.7\%-7.7\%$ by routing strategies (\emph{altruism}).

In order to more effectively mitigate the overall traffic congestion during the Olympics, we proposed an informed \emph{mode change} strategy, in contrast to uninformed practices of restricting cars. To that end, we calculate the contribution of each OD pair to the collective travel time. By reducing the $1\%$ of the total travel demand from the zones near metro and BRT lines, the decrease of overall travel time reaches about $9\%$. Wang \emph{et al.} reported that $1\%$ target decrease in demand can achieve $14\%$ and $18\%$ decrease in travel times for San Francisco Bay Area and Boston Area, respectively~\cite{wang2012understanding}.  However, the proposed countermeasures did not consider the alternative travel modes. In contrast, our strategy only targets drivers within $3km$ of public transportation both in their origins and destination. For incentives, the government could set up discounts for transit and promote ridership services between the selected communities to metro or BRT stations.

Overall we showed that the use of information to target mode change can be the most cost-effective alternative to increasing capacity in transportation. This information-based approach is convenient not only for relieving congestion, but also has the potential favor the use of public transport, deliver better environmental outcomes, stronger communities, and more sustainable cities. We have estimated how the travel demand in each zone contributes differently to the overall congestion, these results can be helpful for the palling of routes of public transportation. In future studies, we can calculate the reduction in emissions associated with the improve in travel times when take one vehicle out from the selected OD pairs. Thereby managing vehicle demand to improve air quality. The data resources used in our work are the byproducts of the use of communication technologies (CDRs, waze, etc) or open source repositories (event schedule, venue property, Airbnb, hotel, OpenStreetMap, etc). Consequently, the proposed methods are portable for events in other cities. Meanwhile, as the data resources are becoming more and more open and abundant, our work represents a very concrete application for demand prediction and management that improves the urban well being.

 Here we evaluated three ideal scenarios and their impact over the Olympics. We expect that the most likely routing behavior to be observed will be between \emph{habit} and \emph{selfish}, meaning that only part of the population may find their shortest route and others will follow their habit route before Olympics. To have an idea of such scenarios, we have defined a selfish parameter $\Lambda$, and report the results for different values that go from habit to selfish case. 
 
 Interesting avenues for improving this work is the estimation of routing behavior~\cite{lima2016understanding} . Collecting data about individual route choices before and after the event will be useful to understand the changes of behavior during large events.

\section{\label{method}Method}

\subsection{Data sets}
The data resources used in this work are: mobile phone data (CDRs), Waze data, camera data, Airbnb data, hotel data, Olympic game schedules and location, as well as the OpenStreetMap. CDRs consist of 5 months of $2.19$ million users and are used to estimate the 24-hour routine ODs before Olympics; Waze data sets contain $0.6$ million reports in one month and are used to extend the 24-hour ODs to five weekdays. We argue that the larger overall congestion length in the road network relates to larger number of cars. Also, camera data sets provided the relation between traffic volumes and average speeds in $85$ main streets and are used to calibrate the relationship between volume-over-capacity and actual travel time. Airbnb data sets contain $13,400$ properties and each property provides its location and the number of accommodations available. We estimate the distribution of tourists' residences using Airbnb data set together with $106$ hotels information. OpenStreetMap provides the road network we used in our demand assignment. Game schedules and locations and capacities of the venues are used for the estimating the tourists' destination and departure times. Among the data sets, CDRs, Waze data, and camera data are the byproduct of the activity. Other data sets are all publicly available (see Supplementary Notes 1-4).

\subsection{Tourists travel mode split}
Before we estimate the vehicle demand during Olympics, the taxi demand of tourists must be split from the tourists demand in each hour. We define four modes for tourists: walking and metro/BRT, bike and metro/BRT, taxi, and bus. The reason we merge the metro line and BRT lines together is that they are closed-loop, as shown in Figure~\ref{fig:map}. Walking and metro/BRT means the origin and destination of tourists is near enough to the stations ($1km$); Bike and metro/BRT means they are near enough for biking ($2km$). Besides, the tourists will consider bus if its travel time and number of transfers are both acceptable. Otherwise, they will choose taxi to the venues. Besides, we assume the occupancy of taxi by tourists is $2.0$, which means two tourists will take one taxi averagely during the Olympics (see Supplementary Figure 2 and Note 3).

\subsection{Travel time estimation}
To estimate travelers' delay during the Olympics, we represent their routing behavior before and during Olympics using a traffic assignment model. Traffic assignment aims to estimate the travel time and volume on each road segment. The estimation is implemented by appointing a reasonable (usually shortest) travel path for all of the trips from their origin to destination. Before Olympics, we assume that all travelers have found their route with shortest travel time and assign the demand with UE model. To validate the estimated travel time, we compare the travel times of top 5000 commuter OD pairs with Google map API during the morning peak hour. The results show the estimation is acceptable (see Supplementary Figure 1). During Olympics, both of the demand and the capacity of the road networks change. For the \emph{habit} scenario, as all of the travelers follow their route before the Olympics, we update the volume and travel time on each edge with considering the additional tourists flow. Tourists' routes are chosen according to the shortest path before Olympics. For the \emph{selfish} scenario, we assign the new demand with UE model as before the Olympics. For \emph{altruism} scenario, we calculate the shortest path with respect to the marginal cost for each OD pair, which makes the entire road network reach system optimum. Aiming at a more realistic estimation of travelers' routing after Olympics starts, we argue that only a fraction of people can find their shortest path, which means one fraction of the drivers follow their routine route, while the remaining fraction is assigned using the UE model to the available space (see Supplementary Figure 4 and Note 5).

\vspace{7mm}
\begin{acknowledgments}
We thank Shan Jiang and Yingxiang Yang for informative discussions and code at the beginning of this project. Armin Akhavan for helpful suggestions with the visualization platform. We specially thank all the members of the Sala Pensa  of Rio de Janeiro's Major Office, lead by Pablo Cerdeira for discussions and data provided. We thank Prof. Alexandre G. Evsukoff for data and contextual information on Rio de Janeiro. The research was partly funded by Ford, the Department of Transportation's grant of the New England UTC Y25, the MIT Portugal Program, the MIT-Brazil seed Grants Program and the Center for Complex Engineering Systems at KACST-MIT.
\end{acknowledgments}



\bibliography{apssamp}

\end{document}